\documentclass[12pt]{article}         
\usepackage{osa2}
\usepackage{overcite}  

\begin{document}

\title{A comparison of designs of off-axis Gregorian telescopes for
mm-wave large focal plane arrays}

\author{Shaul Hanany}
\address{School of Physics and Astronomy\\
  University of Minnesota\\
  116 Church St. SE, Minneapolis, Minnesota, 55455}
\email{hanany@physics.umn.edu}
\author{Daniel P. Marrone}
\address{School of Physics and Astronomy\\
  University of Minnesota\\
  116 Church St. SE, Minneapolis, Minnesota, 55455; 
  Current address: Harvard-Smithsonian Center for Astrophysics\\
  60 Garden Street\\
  Cambridge, Massachusetts, 02138 }

\begin{abstract}
We compare the diffraction-limited field of view (FOV) provided by
 four types of off-axis Gregorian telescopes: the classical Gregorian,
 the aplanatic Gregorian, and designs that cancel astigmatism and both
 astigmatism and coma. The analysis is carried out using telescope parameters
 that are appropriate for satellite and balloon-borne millimeter and
 sub-millimeter wave astrophysics.  We find that the design that cancels
 both coma and astigmatism provides the largest flat FOV, about 21 square
 degrees.  We also find that the FOV can be
 increased by about 15$\%$ by optimizing the shape and location 
 of the focal surface.
 
\end{abstract}


\section{Introduction}
\label{Intro}

Future advances in mm and sub-mm wave astronomy critically depend on
the design of optical systems. Detector technology has matured to the
point where detector sensitivity is limited by photon noise from the
source or from unavoidable photons in the light-path, such as the
atmosphere and the mirrors.  Only an increase in the number of
photometers can significantly increase the overall sensitivity of an
instrument.

Arrays of detectors with tens and hundreds of elements have recently
come on line\cite{scuba,bolocam} and it is widely expected that the
construction of such instruments will accelerate as the fabrication of
both semiconductor-based and superconducting-based bolometers becomes
more uniform and more automated through the use of standard
micro-lithography techniques
\cite{Agnese99,Benford98,Bock98,Kreysa98,Gildemeister00}.

These large focal plane arrays need to be coupled to telescopes that
provide a correspondingly large, diffraction-limited field of view
(FOV). The need for optical systems with large useable FOV was not
acute in the past when typically only a few photometers were coupled
to the telescope. Because of the long wavelength, it is relatively
easy to design an optical system for use at the millimeter wave band
that is diffraction limited near the center of the FOV. The current
challenge is to provide for the {\it largest} useable FOV in order to
accommodate large arrays.

An increase in the available FOV of off-axis Gregorian telescopes is
of particular interest because such telescopes have higher aperture
efficiency and lower side-lobe response than Cassegrain or on-axis
Gregorian reflecting telescopes \cite{DragoneHogg}.  It is also
interesting to analyze in detail Gregorian telescopes that have a low
f-number and a small number of mirrors because such systems find wide
use in satellites and balloon borne payloads which require compact and
simple optical systems.  For example, Gregorian telescopes have been
used extensively in recent years in ground based and balloon borne
experiments to characterize the anisotropy of the cosmic microwave
background radiation (CMB) \cite{archeops,maxima,boomerang,viper}.
Both NASA's Microwave Anisotropy Probe (MAP) satellite and the
European Space Agency's Planck satellite, two missions designed to map
CMB temperature fluctuations, employ off-axis Gregorian telescopes
with low f-number \cite{map,planck}.

The primary source of aberrations in Gregorian and Cassegrain
telescopes are coma and astigmatism. Several designs have been
proposed to improve on the classical Gregorian (CG) design, which has
a parabolic primary and an elliptical secondary.  In an aplanatic
Gregorian (AG) telescope coma is cancelled without creating spherical
aberration. The primary mirror is slightly ellipsoidal and the
secondary is a slightly more eccentric ellipsoid than in the similar
CG. The conic constant of one mirror is chosen to eliminate coma and
the conic constant of the other mirror is adjusted to compensate the
spherical aberration introduced by the change in the first
mirror. Aplanatic designs are fairly common, for example, aplanatic
versions of Cassegrain telescopes were used in the Hubble Space
Telescope \cite{hubble} and the Keck 10-meter telescopes
\cite{keck}.

Dragone \cite{Dragone82,Dragone83} has described designs for off-axis
Gregorian systems which eliminate astigmatism, and both astigmatism
and coma; hereafter we refer to these designs as D1 and D2,
respectively. These designs also greatly reduce instrumental
polarization for field points near the center of the FOV
\cite{Mizuguchi78}. The reduction in instrumental polarization is of
benefit for antennas for communication systems, which use polarization
as a method to increase band-width \cite{Dragone78,Westcott79}, and
for experiments designed for detecting polarized signals. For example,
intense efforts are now being made by a number of experiments to
discover the CMB polarization anisotropy
\cite{polatron,pique,maxipol,polar,compass}.

The AG, D1 and D2 designs present progressively improved image quality
near the center of the field of view, however it is not clear which of
the systems provides a larger useable FOV, which is the quantity of
interest for millimeter wavelength focal plane arrays.  In this paper
we quantitatively compare the size of the diffraction-limited FOV
provided by these three optical designs.  Because we are interested in
potential applications for millimeter-wave astrophysics, and CMB
research in particular, we perform our analysis with telescopes that
provide $\sim 8$ arcminute full-width at half-maximum beam size at 150
GHz ($ \lambda = 2$ mm). We comment on the applicability of our
analysis to other wavelengths in Section 3.

We use the telescope of the Archeops balloon borne
experiment\cite{archeops} as a baseline for comparison. Archeops is
designed to observe the CMB with an array of 24 bolometric photometers
distributed in four frequency bands between 143 and 545 GHz with beam
sizes between 8 and 5 arcminutes, respectively. The focal plane array
is a prototype of the High Frequency Instrument, one of two focal
plane instruments on board ESA's Planck satellite. The satellite is
scheduled to be launched around 2007. Archeops has an off-axis tilted
Gregorian telescope following the D1 design, and is similar in its
physical parameters to the Planck telescope \cite{planck}.

\section{Method of Comparison}

To compare the three optical designs in terms of their useable field
of view we held the focal ratio $f$, aperture $a$, and off-axis
distance of the chief ray $y$ constant with values
as those of the Archeops telescope. (The off-axis 
distance of the chief ray is measured from the primary mirror 
paraboloid axis, see
Figure~\ref{Fig-D1-D2}).  These values are given in
Table~\ref{Table-Archeops}.  The magnitude of coma and astigmatism
aberrations scale as $A_{c}(\theta y^{2} / f^{2})$ and
$A_{a}(\theta^{2} y /f)$ respectively, where $\theta$ is the field
angle and $A_{c}$ and $A_{a}$ are coefficients that depend on other
parameters of the telescope, such as magnification, and mirror conic
constants.  By keeping $f$, $a$, and $y$ constant we ensure that all
three telescopes have the same angular resolution and very similar
physical size, allowing us to examine how the different designs affect
the coefficients $A_{c}$ and $A_{a}$.

We use the diffraction-limited FOV (DLFOV) as the figure of merit by
which we compare the different optical systems. As is customary
\cite{SmithWJ}, we call the image of a field point "diffraction-
limited" if the root mean square wavefront error (WFE) is less than
1/14 of a wave. We measure WFEs on a grid of elevation and
cross-elevation using the optical design software CodeV.  To convert
a set of WFE measurements to a FOV we determine the angular interval
in elevation and cross-elevation at which the image is
diffraction limited. We approximate the diffraction-limited region of
the focal plane as an ellipse with axes defined by these intervals 
and calculate the area of this ellipse as: DLFOV area =
($\pi/4$)[elevation interval$\times$cross-elevation interval]. 
Detailed analysis showed
that in all cases the approximation of the DLFOV as an ellipse was
within 5\% of the DLFOV calculated numerically using a finely spaced grid 
of WFE measurements.  We have chosen the
DLFOV as a figure of merit because it provides a direct measure of the
number of photometers that can be coupled to the telescope.

We compare the three telescopes using two classes of focal surfaces, a
nominal focal surface and an optimized focal surface. The nominal
focal surfaces for the AG, D1, and D2 systems are flat and positioned
at the final focus of each system.  The optimized focal surfaces are
curved and displaced from the position of the nominal surface in order
to provide larger DLFOVs. To determine the optimized focal surface we
evaluate the WFE numerically at 25 locations on the surface as a
function of its radius of curvature, defocus, and tilt\cite{en-Tilt},
and the optimal surface is the one that provides the largest DLFOV. We
choose these three parameters for optimizing the focal surface because
they provide a broad range of uncomplicated focal surface
configurations.  It is possible to further improve telescope
performance by increasing the complexity of the focal surface, for
example, by specifying a position-dependent defocus.  However, such
solutions are too specific to a given photometer-array configuration
to be useful for a comparison of the general properties of telescope
design.

Although the focal surface optimization tends to degrade the image at
the center of the focal plane and to improve the performance at the
edge, the WFE at the center of the focal plane in all three optimized
systems remains far below the diffraction limit.
Figure~\ref{Fig-Fieldpts} shows the 25 field points used for each
optimization\cite{en-CodeV}; all points are weighted equally.  We have
tested many different weighting schemes and concluded that the DLFOV
obtained with each point weighted equally is within one or two percent
of the best DLFOV achieved with more complicated weighting schemes.

\section{Results and Conclusions}
\label{Results}

The results, summarized in Table~\ref{Table-FOV}, show that the D2
design provides the largest available FOV and is a good choice as a
telescope that needs to accommodate a large array of photometers. The
AG design provides the smallest FOV, although it is still considerably
better than the classical Gregorian telescope upon which all of these
systems are based.  With our telescope parameters, the optimized D1
and D2 designs provide $\sim 20\%$ and $\sim 50\%$ larger DLFOV than
the optimized AG design, respectively. As expected, a larger DLFOV is
obtained in all systems by optimizing the parameters of the focal
surface, but this improvement decreases from $\sim 60\%$ in the case
of the AG design to only about 20\% for the D2 design. Because the D2
system with a flat focal plane provides a useable FOV that is almost
as large as the one with an optimized focal surface, it is very
suitable for arrays of detectors that are fabricated on flat silicon
wafers \cite{Agnese99,Benford98,Bock98,Kreysa98,Gildemeister00}. For
this system the physical lengths of the axes of the diffraction
limited region of the focal plane are 17.7 and 16.2 cm in the
elevation and cross-elevation directions, respectively.

The DLFOV that we found for each of the three telescope designs is the
area in which the effects of aberrations are small compared to
diffraction for a frequency of 150 GHz and an aperture that gave a
single mode beam size of 8 arcminutes. It is straight-forward to show
that for a fixed beam size and in the single mode optics limit the
DLFOV will be larger at higher frequencies.  For single mode optics
the aperture area $A$, beam solid angle $\Omega$, and frequency $\nu$
are related through $A\Omega = C/\nu^{2}$, so for a fixed beam size $A
\propto \nu^{-2}$; at higher frequencies the aperture area is smaller.
Since the diffraction spot (or Airy disk) size scales as $1/(A
\nu^{2})$, it is a constant as a function of frequency under these
assumptions. Ray aberrations, however, decrease as $A$ decreases so
the ratio of aberration size to diffraction spot size also decreases
with increasing $\nu$.  Because a single mode system of constant
beamsize becomes increasingly diffraction limited at higher
frequencies, the DLFOV is likely to increase.

Since the D1 and D2 designs can significantly improve the available
area in the FOV it is instructive to assess the shape of the mirrors
that these designs require. The D1 mirrors are conic sections, an
ellipsoid and a paraboloid, identical to those that define the CG, but
the ellipsoid axis is tilted relative to the paraboloid axis. This
tilt is chosen according to conditions outlined by
Dragone\cite{Dragone82}; in the case of the Archeops telescope the
angle is 15$^{\circ}$ (see Table~\ref{Table-Archeops} and
Figure~\ref{Fig-D1-D2}). In the D2 design, localized corrections are
applied to the shape of the mirrors of the D1 design. These
corrections are designed to cancel coma near the center of the
FOV\cite{Dragone83}. The magnitude of the local surface corrections
are given by $ K\, r^{4}$, where $K$ is a constant that is different
for each of the two mirrors and $r$ is the perpendicular distance from
the the segment of the optical axis between the two mirrors, see
Figure~\ref{Fig-D1-D2}.

The constant $K$ depends on the distance between the mirrors along the
optical axis, and for Gregorian telescopes the corrections are such
that they curve the primary toward the secondary and the secondary
away from the primary (see Figure~\ref{Fig-D1-D2}). For the Archeops
system that we have discussed in this paper $K = 3.54\times10^{-9}$
and $7.76 \times 10^{-8}$ cm$^{-3}$ for the primary and secondary,
respectively. Given the sizes of the two Archeops mirrors the largest
correction of the primary is 3.4 mm and is 2.4 mm for the secondary,
values which are neither very large compared to the size of the
mirrors, nor so small such as to make accurate machining difficult.


\newpage
\section*{List of Figures}

\begin{figure}[h]
\caption{The D1 and D2 systems compared in this paper. The
correction applied to the D2 system has been increased by a factor of
25 to make it visible in this plot. Some parameters of the D1 system
given in Table~\ref{Table-Archeops} are labeled.}
\label{Fig-D1-D2}
\end{figure}

\begin{figure}[h]
\caption{The field points used in the focal plane optimizations. The
crosses ($+$) mark the AG field points and the diamonds ($\diamond$)
mark the field points used for both the D1 and D2 designs.}
\label{Fig-Fieldpts}
\end{figure}

\newpage
\centerline{\rotatebox{-90}{\scalebox{.6}{\includegraphics{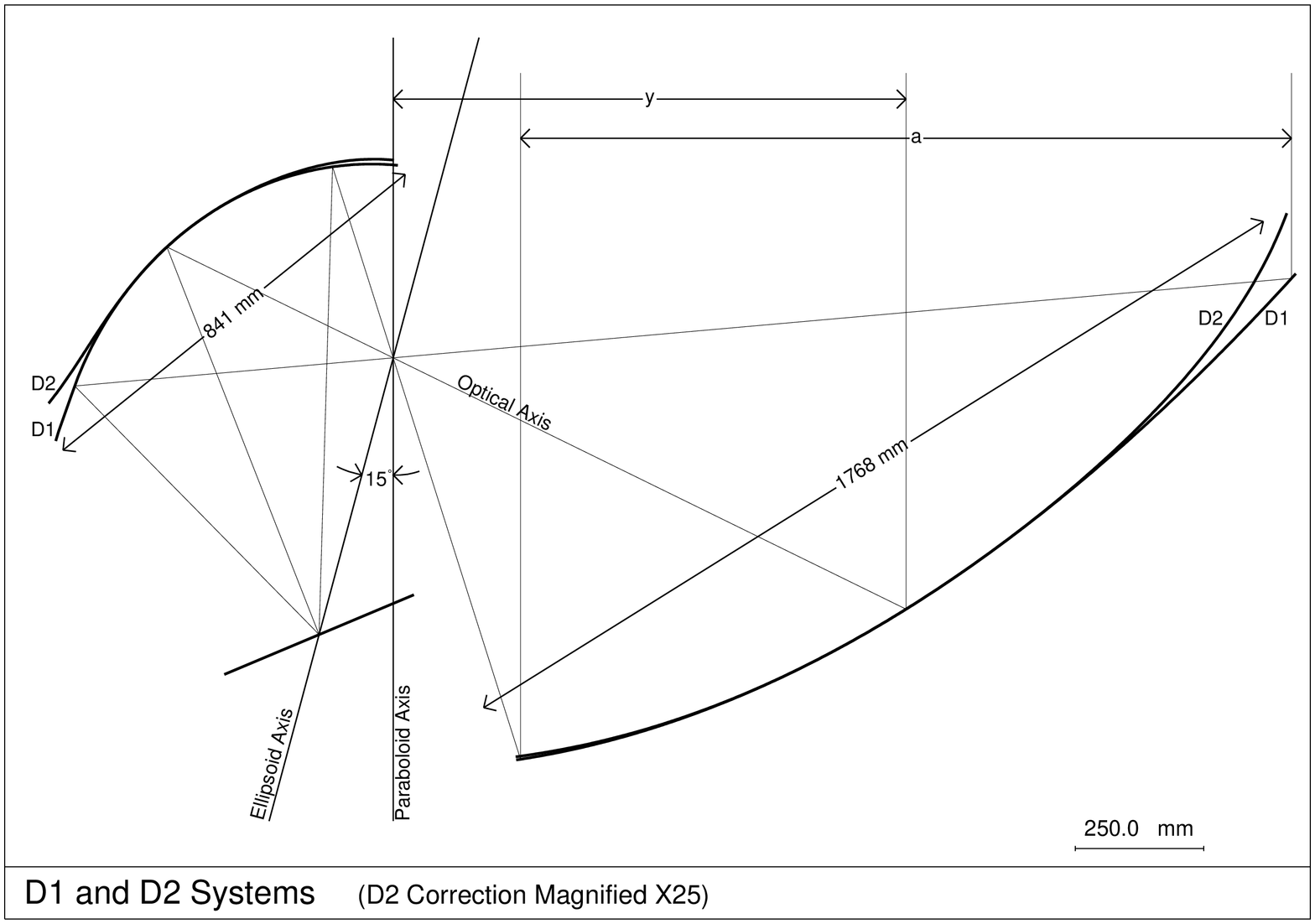}}}}
\vskip2in
Figure 1, S. Hanany and D. P. Marrone.

\newpage
\centerline{\rotatebox{90}{\scalebox{.7}{\includegraphics{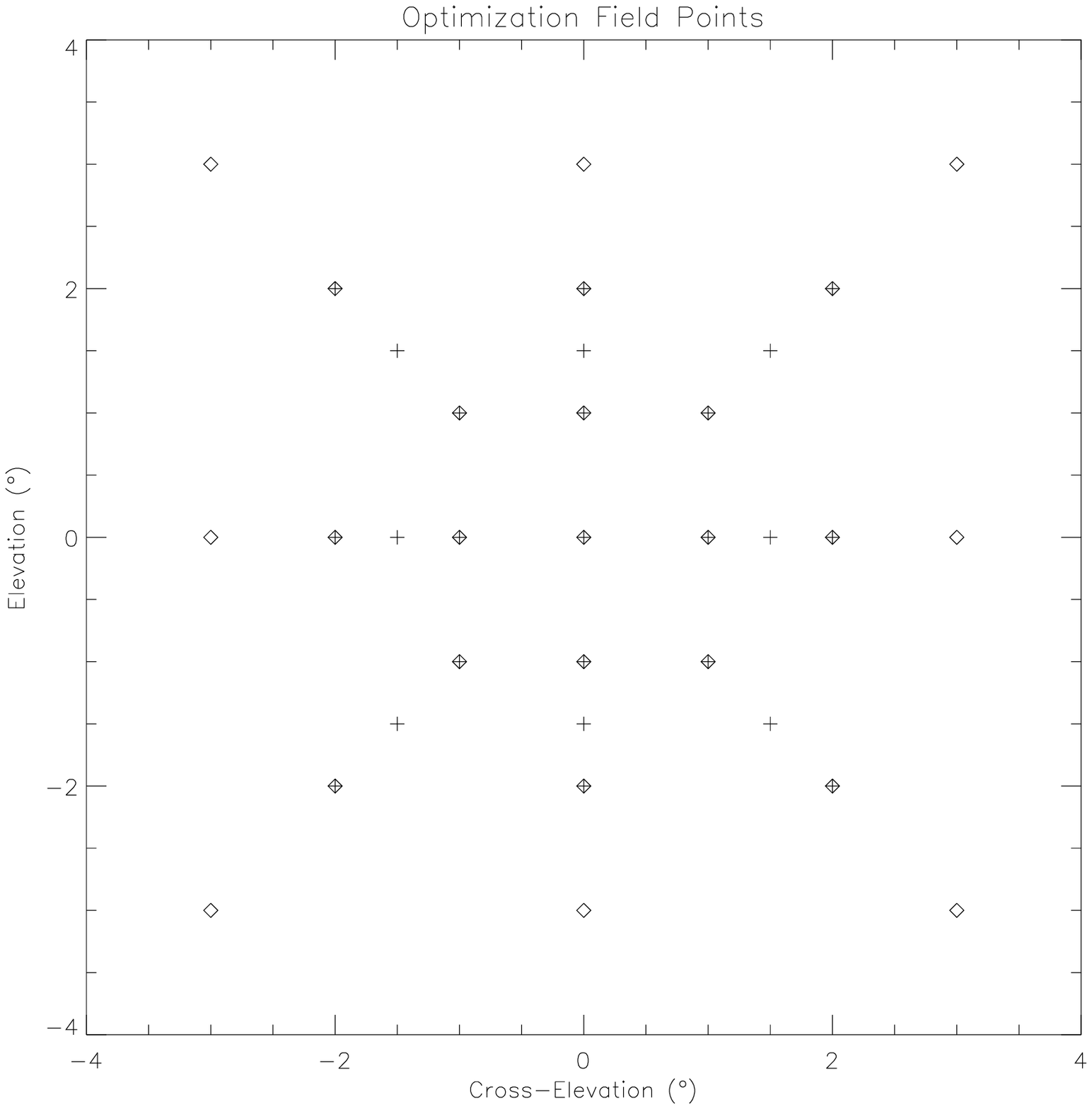}}}}
\vskip2in
Figure 2, S. Hanany and D. P. Marrone.


\begin{table}[h]
\vskip-.7in
\caption{Various parameters of the Archeops telescope.}
\begin{center}
\begin{tabular}{l|r}
\hline
Aperture ({\it a})              & 1500 mm\\
Focal Ratio ({\it f})           & 1.33\\
Off-Axis Distance ({\it y})     & 997.79 mm\\
Relative Tilt of Mirror Axes    & 15$^{\circ}$\\
\hline
Primary Mirror: & \\
Physical Dimensions             & 1500 mm $\times$ 1768 mm\\
Shape                           & Paraboloid\\
Focal Length                    & 800 mm\\
\hline
Secondary Mirror: & \\
Physical Dimensions             & 790 mm $\times$ 841 mm\\
Shape                           & Ellipsoid\\
Semi-major Axis                 & 650 mm\\
Conic Constant                  & -0.1837\\
\hline

\end{tabular} 
\label{Table-Archeops}
\end{center}
\end{table}

\begin{table}
\vskip-.7in
\caption{Elevation and cross-elevation intervals, 
and diffraction limited field of view 
(DLFOV) for various designs of off-axis Gregorian telescopes}
\begin{center}
\begin{tabular}{l|cc|cc}
\hline
System & Elevation       & Cross-Elevation & \multicolumn{2}{c}{DLFOV$^{a}$}   \\
  & (deg.) & (deg.) & (sq. deg.) & ($10^{-3}$ sr.) \\
\hline

Nominal CG$^{b}$ & 1.30 & 2.20 & 2.20 &  0.67 \\
\hline
Nominal AG$^{c}$ & 3.85 & 3.30 & 10.0 & 3.05 \\
Optimized AG & 4.85 & 4.40 & 16.8 & 5.12 \\
\hline
Nominal D1$^{d}$ & 4.95 & 4.60 & 17.9 & 5.45 \\
Optimized D1 & 5.35 & 4.70 & 19.7 & 6.00 \\
\hline
Nominal D2$^{d}$ & 5.45 & 5.00 & 21.4 & 6.52 \\
Optimized D2 & 5.95 & 5.20 & 24.3 & 7.40 \\
\hline 
\multicolumn{5}{l}{$^{a}$ The DLFOV is given in both square degrees and steradians} \\
\multicolumn{5}{l}{$^{b}$ Classical Gregorian} \\
\multicolumn{5}{l}{$^{c}$ Aplanatic Gregorian} \\
\multicolumn{5}{l}{$^{d}$ D1 and D2 are defined in the text} \\ 
\end{tabular} 
\label{Table-FOV}
\end{center}
\end{table}


\begin{thebibliography}{}
\bibitem{en-Tilt} The tilt is about an axis perpendicular to the plane
  of the optical axis.
\bibitem{en-CodeV} Our optimization was performed using the CodeV
  optimization routine, using our own optimization criteria intended
  to maximize DLFOV.
\end{thebibliography}
\end{document}